# Combined estimation for multi-measurements of branching ratio

LIU Xiao-Xia(刘晓霞)[1]    LYU Xiao-Rui(吕晓睿)[1, 3, 1)]
ZHU Yong-Sheng(朱永生)[1, 2, 2)]

[1] University of Chinese Academy of Sciences, Beijng 100049, China

[2] Institute of High Energy Physics, Chinese Academy of Sciences, Beijing 100049,China

[3] CAS Center for Excellence in Particle Physics, Beijing 100049, China

**Abstract:** A maximum likelihood method is used to deal with the combined estimation of multi-measurements of a branching ratio, where each result can be presented as an upper limit. The joint likelihood function is constructed using observed spectra of all measurements and the combined estimate of the branching ratio is obtained by maximizing the joint likelihood function. The Bayesian credible interval, or upper limit of the combined branching ratio, is given in cases both with and without inclusion of systematic error.

## 1. Introduction

Measurements of branching ratios of resonances are essential in high energy physics experiments. Usually, for a particular decay channel of a resonance, different experiments may carry out their respective measurements of its branching ratio. In other cases, a single experiment can implement measurements for the same branching ratio through different decay chains. Combining these results of a branching ratio based on certain statistical methods will usually lead to a better precision than any individual measurement.

Suppose there are $I$ independent measurements of a quantity, their observed values are expressed as $x_i \pm \sigma_i, i=1,\cdots,I$. Assuming the measurements follow the normal distribution, the combined estimate of these $I$ independent measurements for the quantity can be expressed as $\mu \pm \sigma$, where

$$\mu = \frac{\sum_{i=1}^{I}\frac{x_i}{\sigma_i^2}}{\sum_{i=1}^{I}\frac{1}{\sigma_i^2}}, \qquad \sigma = \frac{1}{\sqrt{\sum_{i=1}^{I}\frac{1}{\sigma_i^2}}}. \tag{1}$$

Equation (1) can be used to give a combined estimate of multi-measurements for a branching ratio in the case that each result has the form $x_i \pm \sigma_i$. However, in some cases, which are not rare, some measurements can only report upper limits for a branching ratio, due to the low statistics of the signal events. In such case, Equation (1) is not applicable to deduce the combined estimate of multi-measurements of a branching ratio.

*Supported by National Nature Science Foundation of China (11275266) and Youth Innovation Promotion Association，CAS
1) E-mail: xiaorui@ucas.ac.cn
2) E-mail: zhuys@ihep.ac.cn

In this article we focus our discussion on a particular but often encountered situation in high energy physics experiments. In each experiment, after applying certain selection criteria to the raw data, the data set of the candidate signal events in the signal region is obtained. The candidate signal events contain both signal and background events, which can be separated by fitting the observed spectrum of a kinematic variable in the signal region. The shapes of the signal and background functions of the kinematic variable are usually determined by Monte Carlo simulation or a control sample of the data. Using the number of signal events obtained in the fit, a corresponding branching ratio can be determined.

To illustrate the idea clearly, we take the measurement of the branching ratio of $\psi' \to \eta J/\psi$ in $e^+e^-$ collisions (BES experiment) as an example [1]. The experiment selected $\gamma\gamma e^+e^-$ and $\gamma\gamma\mu^+\mu^-$ candidate events, by constraining the invariant mass of the lepton pair to the $J/\psi$ mass; the $\gamma\gamma$ invariant mass spectra of the two sets of candidate events are shown in Fig. 1. Here, the kinematic variable is the $\gamma\gamma$ invariant mass $M_{\gamma\gamma}$, the events inside the peak area at 548 MeV correspond to the $\psi' \to \eta J/\psi$ signal, while the broad smooth distribution corresponds to background. If both shapes of the signal and background functions are known, then by fitting the observed $\gamma\gamma$ invariant mass spectrum, the number of signal events can be determined. The branching ratio of $R \to X$ is calculated by

$$B(R \to X) = \frac{N_s(R \to X \to Y)}{N_R \varepsilon(R \to X \to Y) \cdot BR(X \to Y)}. \qquad (2)$$

In the above measurements, $R$ denotes $\psi'$, $X$ denotes $\eta J/\psi$, $Y$ denotes $\gamma\gamma e^+e^-$ or $\gamma\gamma\mu^+\mu^-$, $N_s$ is the number of observed signal events, $N_R$ is the total number of decays of resonance $R$, and $\varepsilon$ is the detection efficiency of a signal event. The measured value of $\psi' \to \eta J/\psi$ decay branching ratio from $\gamma\gamma e^+e^-$ and $\gamma\gamma\mu^+\mu^-$ channels can be written as (using simplified symbols):

$$B_i = \frac{N_{is}}{N_{Ri}\varepsilon_i BR_i} \equiv \frac{N_{is}}{A_i}, \quad i = 1(\gamma\gamma e^+e^-), 2(\gamma\gamma\mu^+\mu^-), \qquad (3)$$

where the symbol with subscript $i$ represents the value of the $i^{\text{th}}$ measurement. Equation (3) is easy to extend to the cases of $i > 2$. The uncertainty of the measured $B_i$ due to the statistical fluctuation of $N_{is}$ is usually considered as the statistical error of $B_i$, while the uncertainty of $B_i$ due to the uncertainty of $A_i$ is considered as the systematic error.

By fitting the observed spectra shown in Fig. 1a and 1b, the number of signal events $N_{is}$ can be obtained and the measured branching ratio of $\psi' \to \eta J/\psi$ from $\gamma\gamma e^+e^-$ and $\gamma\gamma\mu^+\mu^-$ channels can be calculated with Equation (3) to be $2.91\pm0.12$ and $3.06\pm0.14$ (statistical error only), respectively. Assuming these two measurements are independent, the combined estimate for the branching ratio of $\psi' \to \eta J/\psi$ can be determined with Equation (1). However, if one of the results is an upper limit, Equation (1) is not applicable anymore.

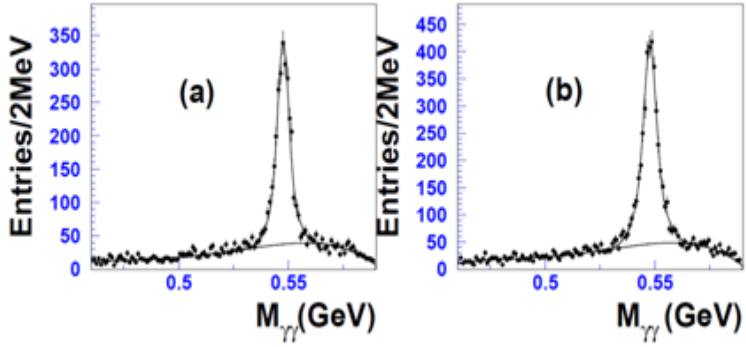

Fig. 1  $\gamma\gamma$ invariant mass spectrum of $\psi' \to \eta J/\psi$ candidate events,

(a) $\gamma\gamma e^+e^-$ channel,  (b) $\gamma\gamma\mu^+\mu^-$ channel.

It is typical to obtain the number of signal events by fitting the observed spectrum of a kinematic variable ($M_{\gamma\gamma}$ in the above example) in branching ratio or cross section measurements. In this article, we describe in detail the maximum likelihood method to deal with the combined estimation for multi-measurements of a branching ratio in cases where some or all the results of these measurements are given as upper limits. This method constructs a joint likelihood function using all observed spectra obtained in individual measurements, maximizes the joint likelihood function, and then obtains the combined estimate of the branching ratio. The basics of the maximum likelihood method can be seen in many text books and references [2,3].

The way to construct a joint likelihood function depends on the forms of observed spectra in individual measurements. In Sections 2 and 3, we describe the combined estimation for individual observed spectra as function of a same kinematic variable and different kinematic variables, respectively. The determination of the Bayesian credible interval and upper limit with and without inclusion of systematic error for the combined branching ratio is discussed in Section 4. The results of a test with Toy Monte Carlo samples are shown in Section 5. Finally, a conclusion is given in Section 6.

## 2. Combined estimation for individual observed spectra as function of a same kinematic variable

**2.1 Individual observed spectra are histograms with same binning**

Suppose there are *I* experiments which measure the same branching ratio of a resonance, each experiment giving an observed spectrum of the candidate signal events in the same signal region as a histogram for a kinematic variable *m* with the same binning.

In this case, a merged spectrum of *I* experiments for the variable *m* can be constructed, whose histogram has the same binning as the individual histograms. Let the number of events in bin *j* for the $i^{th}$ experiment be $n_{ij}, i = 1, \cdots, I, j = 1, \cdots, J$; the number of events in bin *j* is

$$n_j = \sum_{i=1}^{I} n_{ij}, \quad j = 1, \cdots, J. \tag{4}$$

The number of events in the $i^{th}$ experiment is $N_i = \sum_{j=1}^{J} n_{ij}$. The total number of events of the *I* experiments, namely, the total number of events of the merged spectrum is $N = \sum_{i=1}^{I} N_i$. In Equation (4), $n_j$ can be considered as a Poisson variable with expectation $\lambda_j$. The joint likelihood function of observing $n_j$ events in bin *j* ( $j = 1, \cdots, J$ ) is

$$L(n_1, \cdots, n_J) = \prod_{j=1}^{J} \frac{1}{n_j!} \lambda_j^{n_j} e^{-\lambda_j}, \tag{5}$$

where $\lambda_j$ is calculated by the integral of the combined probability density function (pdf) of *I* experiments, $f(m|\boldsymbol{\theta})$, in bin *j*:

$$\lambda_j = \lambda \int_{\Delta m_j} f(m|\boldsymbol{\theta}) dm. \tag{6}$$

Here, *m* is the kinematic variable, $\Delta m_j$ is the range for *m* in bin *j*, $\boldsymbol{\theta}$ is the parameters in the joint likelihood function which is defined by Equation (13), and $\lambda$ is the expectation of the total number of events *N* (Poisson variable):

$$\lambda = \sum_{j=1}^{J} \lambda_j. \tag{7}$$

Let $f_{is}(m|\boldsymbol{\theta}_{is})$ and $f_{ib}(m|\boldsymbol{\theta}_{ib})$ be the pdf of signal and background distributions in the signal region of the $i^{th}$ experiment, respectively. The combined pdf of the merged spectrum, $f(m|\boldsymbol{\theta})$, can be expressed as

$$f(m|\boldsymbol{\theta}) = \sum_{i=1}^{I} \frac{N_i}{N} \left[ w_{is} f_{is}(m|\boldsymbol{\theta}_{is}) + (1 - w_{is}) f_{ib}(m|\boldsymbol{\theta}_{ib}) \right], \tag{8}$$

Here, the function forms of $f_{is}(m|\boldsymbol{\theta}_{is})$ and $f_{ib}(m|\boldsymbol{\theta}_{ib})$ should already be determined for the $i^{th}$ experiment. $\theta_{ib}$ and $\theta_{is}$ are parameters of the background and signal pdf, whose values need to be determined in the combined fit. For instance, in the measurement of the $\psi' \to \eta J/\psi$ decay branching ratio stated above, the variable $m$ is the invariant mass $M_{\gamma\gamma}$, $\theta_{ib}$ can be the coefficients of the polynomial describing background, $\theta_{is}$ can be the central mass and width of the Breit-Wigner function describing the mass distribution of the resonance $\eta$, and the Gaussian resolution of the detector for the invariant mass. For the combined estimation of multi-measurements for a branching ratio, the central mass and width of the Breit-Wigner function for resonance $\eta$ should be identical, while the Gaussian resolution of the detector for the variable $m$ can be different in each experiment. $w_{is}$ is the ratio of the signal events to the total observed events in the signal region for the $i^{\text{th}}$ experiment. That is, the number of signal events can be written as $N_{is} = w_{is} N_i$. The total number of signal events in the merged spectrum is

$$N_s = \sum_{i=1}^{I} N_{is} = \sum_{i=1}^{I} w_{is} N_i.$$

From Equation (3), we have $N_{is} = A_i B_i$, wherein

$$A_i \equiv N_{Ri} \varepsilon_i BR_i, \quad i = 1, \cdots, I. \tag{9}$$

When we implement a combined estimation for a branching ratio, obviously it assumes $B = B_i$. Since $N_{is} = w_{is} N_i$, we have $w_{is} = A_i B / N_i$. Substituting this relation into Equation (8), we get an expression of the combined pdf for the merged spectrum as follows:

$$f(m|\theta) = \sum_{i=1}^{I} \frac{N_i}{N} \left[ \frac{A_i}{N_i} B \cdot f_{is}(m|\theta_{is}) + \left(1 - \frac{A_i}{N_i} B\right) f_{ib}(m|\theta_{ib}) \right], \tag{10}$$

Based on Equation (5), when one omits quantities which are not related to the interested parameters $\theta$, the log-likelihood function is expressed as

$$\ln L = \sum_{j=1}^{J} n_j \ln \lambda_j - \lambda. \tag{11}$$

The likelihood equation is

$$\left. \frac{\partial \ln L}{\partial \theta} \right|_{\theta=\hat{\theta}} = \frac{\partial}{\partial \theta} \left[ \sum_{j=1}^{J} n_j \ln \lambda_j - \lambda \right]_{\theta=\hat{\theta}} = 0. \tag{12}$$

The parameters $\theta$ in the joint likelihood function contain $\theta_{is}, \theta_{ib}$ and $\lambda, B$:

$$\theta = \{B, \lambda, \theta_s, \theta_b\}, \quad \theta_s = \{\theta_{1s}, \cdots, \theta_{Is}\}, \quad \theta_b = \{\theta_{1b}, \cdots, \theta_{Ib}\}. \tag{13}$$

Using any optimization package to solve the likelihood equation (12), one obtains the estimates

$\hat{\theta}$ for parameters $\theta$ (including the combined estimate $\hat{B}$), and their fitting (statistical) errors (including error of $B$, $\sigma_{B,st}$). In an iterative procedure of the maximum $\ln L$ calculation, the initial value of $\lambda$ can be taken as $N$, the initial value of $B$ can be the weighted average of all individual results $B_i$, while the initial values of $\theta_{is}$ and $\theta_{ib}$ use the resultant values obtained in each individual experiment.

If all the background pdfs $f_{ib}(m|\theta_{ib})$ in each experiment are smooth distributions, it is possible to form a merged background pdf $f_b(m|\theta_b)$ in constructing the merged spectrum, namely :

$$f(m|\theta) = \sum_{i=1}^{I} c_i \left[ \frac{A_i}{N_i} B \cdot f_{is}(m|\theta_{is}) \right] + q \cdot f_b(m|\theta_b). \tag{10a}$$

Here, the function form of $f_b(m|\theta_b)$ can be determined empirically according to the shape of the background in the merged spectrum and $q$ represents the ratio of the background events to the total events in the merged spectrum. The parameters $\theta$ in the joint likelihood function contain $\theta_s$, $\theta_b$, $\lambda$, $q$ and $B$ :

$$\theta = \{B, \lambda, q, \theta_s, \theta_b\}, \quad \theta_s = \{\theta_{1s}, \cdots, \theta_{Is}\}. \tag{13a}$$

The remainder of the procedure is the same as before, except Equation (10) is replaced by Equation (10a) and Equation (13) replaced by Equation (13a).

**2.2 Individual observed spectra are unbinned data within the same signal region**

Suppose $I$ experiments measure a branching ratio of a resonance, and each experiment gives an observed spectrum of the candidate signal events in the same signal region as unbinned data for a variable $m$.

Let the number of events in the one-dimensional scatter plot (a collection of points for variable $m$ of a data set) at $i^{\text{th}}$ experiment be $N_i$, these $N_i$ events appear at $m = m_{i1}, \cdots, m_{iN_i}$, $i = 1, \cdots, I$. The total number of events for the merged spectra of $I$ experiments is $N \equiv \sum_{i=1}^{I} N_i$ and the combined pdf in signal region, $f(m|\theta)$ can still be described by Equation (10).

We define the joint likelihood function for these $N$ events as :

$$\begin{aligned} L &= L\left(m_{11}, \cdots, m_{1N_1}; \cdots; m_{I1}, \cdots, m_{IN_I} \mid B, \theta_s, \theta_b\right) \\ &= \prod_{i=1}^{I} \prod_{j=1}^{N_i} \left\{ \sum_{i=1}^{I} \frac{N_i}{N} \left[ \frac{A_i}{N_i} B \cdot f_{is}(m_{ij}|\theta_{is}) + \left(1 - \frac{A_i}{N_i} B\right) f_{ib}(m_{ij}|\theta_{ib}) \right] \right\}. \end{aligned} \tag{14}$$

Then we have

$$\ln L = \sum_{i=1}^{I} \sum_{j=1}^{N_i} \ln \left\{ \sum_{i=1}^{I} \frac{N_i}{N} \left[ \frac{A_i}{N_i} B \cdot f_{is}(m_{ij}|\theta_{is}) + \left(1 - \frac{A_i}{N_i} B\right) f_{ib}(m_{ij}|\theta_{ib}) \right] \right\}. \tag{15}$$

The parameters $\theta$ in the joint likelihood function contain $\theta_s$ and $\theta_b$ and $B$ :

$$\theta = \{B, \theta_s, \theta_b\} \quad \theta_s = \{\theta_{1s}, \cdots, \theta_{Is}\}, \quad \theta_b = \{\theta_{1b}, \cdots, \theta_{Ib}\}. \tag{16}$$

If all the background pdfs $f_{ib}(m|\theta_{ib})$ in each experiment are smooth distributions, the pdf of the merged spectra, $f(m|\theta)$, is represented by Equation (10a). Hence we have

$$\begin{aligned} L &= L\left(m_{11}, \cdots, m_{1N_1}; \cdots; m_{I1}, \cdots, m_{IN_I} \mid B, \theta_s, \theta_b\right) \\ &= \prod_{i=1}^{I} \prod_{j=1}^{N_i} \left\{ \sum_{i=1}^{I} \left[ \frac{A_i}{N} B \cdot f_{is}(m_{ij}|\theta_{is}) \right] + q \cdot f_b(m_{ij}|\theta_b) \right\}, \end{aligned} \tag{14a}$$

and

$$\ln L = \sum_{i=1}^{I} \sum_{j=1}^{N_i} \ln \left\{ \sum_{i=1}^{I} \left[ \frac{A_i}{N} B \cdot f_{is}(m_{ij}|\theta_{is}) \right] + q \cdot f_b(m_{ij}|\theta_b) \right\}. \tag{15a}$$

The determined parameters $\theta$ in the joint likelihood function contain $\theta_s$, $\theta_b$, $q$ and $B$ :

$$\theta = \{B, q, \theta_s, \theta_b\}, \quad \theta_s = \{\theta_{1s}, \cdots, \theta_{Is}\}. \tag{16a}$$

Using any optimization package to solve the maximum $\ln L$ calculation, one obtains the estimates $\hat{\theta}$ for parameters $\theta$ (including the combined estimate $\hat{B}$), and their fitting (statistical) errors (including the error of $B$, $\sigma_{B,st}$).

## 3. Combined estimation for individual observed spectra as a function of different kinematic variables

The combined estimation methods of branching ratio described in Section 2 are applicable merely for the case that all *I* experiments give observed spectra for a same kinematic variable *m*, and their signal regions are the same. In this case, a merged spectrum of *I* experiments for the variable *m* can be constructed, and the corresponding combined estimation method is called the merged spectrum method. In this section, we will discuss the combined estimation methods of the branching ratio in more general cases. That is, all (or part of) *I* experiments give observed spectra for different kinematic variables, and their signal regions can be different or the same. In this case, a merged spectrum of *I* experiments cannot be constructed, hence, the merged spectrum method is not applicable; instead, a simultaneous fit for the observed spectrum in each experiment has to be carried out. However, the equations deduced in such general cases are also applicable for the cases that all *I* experiments give observed spectra for a same kinematic variable *m*, which have different (or the same) signal regions.

### 3.1 Individual observed spectra are histograms with different binning

Suppose *I* experiments measure the same branching ratio of a resonance, and each experiment gives an observed spectrum of the candidate signal events in a signal region as a histogram for a kinematic variable $m_i$. The variable $m_i$, the histogram binning and the signal region in each experiment for these *I* measurements can be different.

Let $N_i$ be the number of events in the signal region for the $i^{th}$ experiment. The histogram of the $i^{th}$ experiment contains $J_i$ bins, and the observed number of events in bin $j_i(=1,\cdots,J_i)$ is $n_{ij_i}$, which is a Poisson variable with the expectation $\lambda_{ij_i}$. The joint likelihood function of observing $n_{ij_i}$ events in bin $j_i$ ($j_i = 1, \cdots, J_i$) is

$$L_i(n_{i1}, \cdots, n_{iJ_i}) = \prod_{j_i=1}^{J_i} \frac{1}{n_{ij_i}!} \lambda_{ij_i}^{n_{ij_i}} e^{-\lambda_{ij_i}}, \tag{17}$$

where $\lambda_{ij_i}$ is calculated by the integral of the pdf $f(m_i | \theta_i)$ at bin $j_i$:

$$\lambda_{ij_i} = \lambda_i \int_{\Delta m_{j_i}} f_i(m_i | \theta_i) dm_i, \tag{18}$$

Here $\lambda_i$ is the expectation of the number of events $N_i$ (Poisson variable):

$$\lambda_i = \sum_{j_i=1}^{J_i} \lambda_{ij_i}. \tag{19}$$

$f_i(m_i | \theta_i)$ is the pdf of variable $m_i$ in the signal region of the $i^{\text{th}}$ experiment:

$$f_i(m_i | \theta_i) = w_{is} f_{is}(m_i | \theta_{is}) + (1 - w_{is}) f_{ib}(m_i | \theta_{ib}). \tag{20}$$

$\theta_{is}$ and $\theta_{ib}$ are parameters describing the signal and background functions respectively in the signal region, and $w_{is}$ is the ratio of signal events $N_{is}$ to total events $N_i$ within the signal region, i.e. $N_{is} = w_{is} N_i$. The function forms of $f_{is}$ and $f_{ib}$ should already be known from the data analysis of the $i^{\text{th}}$ experiment.

From Equation (3), it is known that $N_{is} = A_i B_i$, $i = 1, \cdots, I$. When a combined estimation for a branching ratio is implemented, we straightforwardly assume $B = B_i$. Since $N_{is} = w_{is} N_i$, therefore, $w_{is} = A_i B / N_i$. Substituting this relation into Equation (20), we get:

$$f_i(m_i | \theta_i) = \frac{A_i}{N_i} B \cdot f_{is}(m_i | \theta_{is}) + \left(1 - \frac{A_i}{N_i} B\right) f_{ib}(m_i | \theta_{ib}). \tag{21}$$

The joint likelihood function for $I$ experiments is

$$L = \prod_{i=1}^{I} L_i. \tag{22}$$

Omitting quantities not related to the parameters $\theta$, the log-likelihood function is expressed as

$$\ln L = \sum_{i=1}^{I} \left[ \sum_{j_i=1}^{J_i} \left(n_{ij_i} \ln \lambda_{ij_i}\right) - \lambda_i \right] \tag{23}$$

The parameters $\theta$ in the joint likelihood function contain $\lambda$, $\theta_s$, $\theta_b$ and $B$:

$$\theta = \{B, \lambda, \theta_s, \theta_b\}, \lambda = \{\lambda_1, \cdots, \lambda_I\}, \theta_s = \{\theta_{1s}, \cdots, \theta_{Is}\}, \theta_b = \{\theta_{1b}, \cdots, \theta_{Ib}\}. \tag{24}$$

Using any optimization package to solve the maximum $\ln L$ calculation, one obtains the estimates $\hat{\theta}$ for parameters $\theta$ (including the combined estimate $\hat{B}$), and their fitting (statistical) errors (including error of B, $\sigma_{B,st}$). In an iterative procedure of the maximum $\ln L$ calculation, the initial value of $\lambda = \{\lambda_1, \cdots, \lambda_I\}$ can be taken as $\{N_1, \cdots, N_I\}$, the initial value of B can be the weighted average of all individual results $B_i$, while the initial values of $\theta_s$ and $\theta_b$ use the resultant values from each individual experiment.

**3.2 Individual observed spectra are unbinned data within different signal regions**

Suppose I experiments measure a branching ratio of a resonance, and each experiment gives an observed spectrum of the candidate signal events in different signal regions as unbinned data for the variable $m_i$.

Let the number of events in the scatter plot of the $i^{th}$ experiment be $N_i$, these $N_i$ events appeared at $m_i = m_{i1}, \cdots, m_{iN_i}$, $i = 1, \cdots, I$. The pdf in the signal region for the $i^{th}$ experiment, $f_i(m_i | \theta_i)$, can still be described by Equation (21). The total number of events for the spectra of I experiments is $N \equiv \sum_{i=1}^{I} N_i$.

Defining the joint likelihood function for these N events as:

$$L = L(m_{11}, \cdots, m_{1N_1}; \cdots; m_{I1}, \cdots, m_{IN_I} | B, \theta_s, \theta_b) = \prod_{i=1}^{I} L_i(m_{i1}, \cdots, m_{iN_i} | B, \theta_{is}, \theta_{ib})$$
$$= \prod_{i=1}^{I} \prod_{j=1}^{N_i} \left[ \frac{A_i}{N_i} B \cdot f_{is}(m_{ij} | \theta_{is}) + \left(1 - \frac{A_i}{N_i} B\right) f_{ib}(m_{ij} | \theta_{ib}) \right]. \tag{25}$$

We then have

$$\ln L = \sum_{i=1}^{I} \sum_{j=1}^{N_i} \ln \left[ \frac{A_i}{N_i} B \cdot f_{is}(m_{ij} | \theta_{is}) + \left(1 - \frac{A_i}{N_i} B\right) f_{ib}(m_{ij} | \theta_{ib}) \right]. \tag{26}$$

The parameters $\theta$ in the joint likelihood function contain $\theta_s$, $\theta_b$ and $B$:

$$\theta = \{B, \theta_s, \theta_b\}, \quad \theta_s = \{\theta_{1s}, \cdots, \theta_{Is}\}, \quad \theta_b = \{\theta_{1b}, \cdots, \theta_{Ib}\}. \tag{27}$$

Using any optimization package to solve the maximum $\ln L$ calculation, one obtains the estimates $\hat{\theta}$ for parameters $\theta$ (including the combined estimate $\hat{B}$), and their fitting (statistical) errors (including error of B, $\sigma_{B,st}$).

## 4. Determination of the credible interval and upper limit with or without inclusion of systematic error

**4.1 Without inclusion of systematic error**

Now, we have the estimates $\hat{\theta}$ for parameters $\theta$ (including the combined estimate $\hat{B}$), and their fitting (statistical) errors (including error of B, $\sigma_{B,st}$). The problem we face is how to report our combined branching ratio, namely, report a CL = 68.27% interval or CL = 90% upper limit? To answer this question, an additional flip-flopping policy [4] is needed. For instance, a frequently used flip-flopping policy is that if $B \geq 3\sigma_{B,st}$, we report $B \pm \sigma_{B,st}$ as a CL = 68.27% interval; otherwise, a CL = 90% upper limit $B_{up}$ will be given.

Below, we will use Bayesian Highest Posterior Density (HPD) [2] to perform the interval estimation for the branching ratio $B$. We intend to use a flip-flopping policy based on the Bayesian posterior density.

Given a credible level $CL = \gamma$, the optimal interval in Bayesian statistics is the HPD interval. Let $h(B|\boldsymbol{n})$ be the posterior density for parameter $B$, and $\boldsymbol{n}$ be the observed sample values. Then, the HPD interval for $B$ at a credible level $CL = \gamma$ is $R$, which satisfies

$$P(B \in R | \boldsymbol{n}) = \int_R h(B|\boldsymbol{n})dB = \gamma. \qquad (28)$$

and for any $B_1 \in R, B_2 \notin R$, the following relation holds:

$$h(B_1|\boldsymbol{n}) \geq h(B_2|\boldsymbol{n}). \qquad (29)$$

The upper limit $B_{up}$ at $CL = \gamma$ is

$$P(B \leq B_{up}|\boldsymbol{n}) = \int_{B \leq B_{up}} h(B|\boldsymbol{n})dB = \gamma, \qquad (30)$$

For the case of parameter $B$ being the combined branching ratio, $\boldsymbol{n}$ represents the observed spectra of $I$ experiments. For the following three types of observed spectra (a) histograms with the same binning for a same kinematic variable, (b) histograms with different binning for different kinematic variables and (c) unbinned data, from Equations (5), (17), (14), (14a) and (25), $\boldsymbol{n}$ can be expressed as

$$\begin{aligned} \boldsymbol{n} &= \{n_1, \cdots, n_J\}, \\ \boldsymbol{n} &= \{n_{11}, \cdots, n_{1J_1}; \cdots; n_{I1}, \cdots, n_{IJ_I}\}, \\ \boldsymbol{n} &= \{m_{11}, \cdots, m_{1N_1}; \cdots; m_{I1}, \cdots, m_{IN_I}\}. \end{aligned} \qquad (31)$$

$h(B|\boldsymbol{n})$ is the posterior pdf for $B$:

$$h(B|\boldsymbol{n}) = \frac{L(\boldsymbol{n}|B)\pi(B)}{\int L(\boldsymbol{n}|B)\pi(B)dB}. \qquad (32)$$

Here, $L(\boldsymbol{n}|B)$ can be calculated based on Equations (5), (14) or (14a), (22) and (25) for a given $B$, with all other parameters in $\boldsymbol{\theta}$ taken as the values where the likelihood function reaches its maximum. $\pi(B)$ is the prior pdf for $B$, for which we use the flat distribution in the allowed region of $B$[0,1]. It leads to

$$h(B|\boldsymbol{n}) = \frac{L(\boldsymbol{n}|B)}{\int_0^1 L(\boldsymbol{n}|B)dB}. \qquad (33)$$

We use the following flip-flopping policy to decide how to report our combined branching ratio, namely, to report a $CL = 68.27\%$ interval or $CL = 90\%$ upper limit. If there exists a HPD interval at $CL = 90\%$, $R_{0.9}$, and it satisfies

$$R_{0.9} \in [\tilde{B}_l, \tilde{B}_u], \quad \tilde{B}_l < \tilde{B}_u, \quad L(\boldsymbol{n}|\tilde{B}_l) = L(\boldsymbol{n}|\tilde{B}_u), \quad \tilde{B}_l \in [0,1], \quad \tilde{B}_u \in [0,1]. \qquad (34)$$

Then a CL=68.27% interval $R_{0.6827}$ is reported as the measured value of the combined branching ratio:

$$R_{0.6827} \in [B_l, B_u], \quad B_l < B_u, \quad L(\mathbf{n}|B_l) = L(\mathbf{n}|B_u), \quad B_l \in [0,1], \quad B_u \in [0,1] \tag{35}$$

which corresponds to

$$B = \hat{B}^{+\sigma_+}_{-\sigma_-}, \quad \sigma_+ = B_u - \hat{B}, \quad \sigma_- = \hat{B} - B_l, \tag{36}$$

Here $\hat{B}$ is the maximum likelihood estimate of $B$. If $R_{0.9}$ does not exist, we report the upper limit $B_{up}$ at CL = 90% according to Equation (30).

**4.2 Inclusion of systematic error**

In order to estimate the systematic error of the combined estimate of $B$, it is necessary to take into account the correlation between each experiment.

If $I$ measurements for a branching ratio $B$ are independent, the systematic error of $B$ can be calculated by the following equations:

$$\sigma^2_{B,sys} = \left(\sum_{i=1}^{I} \sigma^{-2}_{B_i,sys}\right)^{-1}. \tag{37}$$

$$\frac{\sigma^2_{B_i,sys}}{B_i^2} = \frac{\sigma^2_{N_{Ri}}}{N_{Ri}^2} + \frac{\sigma^2_{BR_i}}{BR_i^2} + \frac{\sigma^2_{\varepsilon_i}}{\varepsilon_i^2} + \frac{\sigma^2_{N_{ib}}}{N_{ib}^2}, \tag{38}$$

where $\sigma_{B_i,sys}, \sigma_{\varepsilon_i}, \sigma_{N_{ib}}$ are the systematic error for $B_i, \varepsilon_i, N_{ib}$ in the $i^{th}$ experiment, respectively. $N_{ib}$ is the expectation of the number of background events in the observed spectrum for the $i^{th}$ experiment: $N_{ib} = N_i - A_i B_i$. All the quantities on the right side of Equation (38) should already be determined from the $i^{th}$ experiment data analyses. By nature, $\sigma_{N_{Ri}}, \sigma_{BR_i}, \sigma_{\varepsilon_i}, \sigma_{N_{ib}}$ are independent of each other.

If $I$ measurements for a branching ratio $B$ are not independent, there is an independent component $\left(\sigma^2_{B_i,sys}\right)_{uncom}$ and a common component $\left(\sigma^2_{B,sys}\right)_{com} \equiv \left(\sigma^2_{B_i,sys}\right)_{com}$ in $\sigma^2_{B_i,sys}$. Then the systematic error for $B$ can be expressed as

$$\sigma^2_{B,sys} = \left(\sigma^2_{B,sys}\right)_{uncom} + \left(\sigma^2_{B,sys}\right)_{com}, \tag{39}$$

$$(\sigma^2_{B,sys})_{uncom} = (\sum_{i=1}^{I} (\sigma^{-2}_{B_i,sys})_{uncom})^{-1}. \tag{40}$$

For instance, in the measurement of the branching ratio of $\psi' \to \eta J/\psi$ stated above, $\sigma^2_{N_{Ri}}$ is the common component, while $\sigma_{BR_i}, \sigma_{\varepsilon_i}, \sigma_{N_{ib}}$ are independent components.

The total error of $B$ is

$$\sigma_B^2 = \sigma_{B,st}^2 + \sigma_{B,sys}^2 \quad . \tag{41}$$

In the case of inclusion of the systematic error of $B$, the likelihood function may depend on the parameter of interest $B$ as well as on a nuisance parameter $\nu$, which is the observed value of the branching ratio off center from $B$ and must be included for an accurate description of the data[2]. Thus, the likelihood function depends on both $B$ and $\nu$, written as $L(\boldsymbol{n}|B,\nu)$. One might characterize the uncertainty in a nuisance parameter $\nu$ by a pdf $\pi(\nu)$ centered about its nominal value with a certain standard deviation $\sigma_\nu$. Here we take the systematic error of the combined branching ratio equal to the error of $\nu$, i.e. $\sigma_{\nu,sys}$. Thus, it can be written as

$$L(\boldsymbol{n}|B,\nu) = L(\boldsymbol{n}|\nu) \cdot \frac{1}{\sqrt{2\pi}\sigma_{\nu,sys}} \exp\left(-\frac{(\nu-B)^2}{2\sigma_{\nu,sys}^2}\right). \tag{42}$$

In this case, the likelihood function $L(\boldsymbol{n}|B)$ in Equation (33) has to be replaced by

$$\tilde{L}(\boldsymbol{n}|B) = \int_0^1 \left[ L(\boldsymbol{n}|\nu) \cdot \frac{1}{\sqrt{2\pi}\sigma_{\nu,sys}} \exp\left(-\frac{(\nu-B)^2}{2\sigma_{\nu,sys}^2}\right) \right] d\nu. \tag{43}$$

Therefore, in the case of inclusion of systematic error, we still use Equations (28~29) and Equations (33~36) to determine the CL = 68.27% interval and use Equations (30) and (33) to determine the CL = 90% upper limit for combined branching ratio $B$, with the posterior density $h(B|\boldsymbol{n})$ of

$$h(B|\boldsymbol{n}) = \frac{\tilde{L}(\boldsymbol{n}|B)}{\int_0^1 \tilde{L}(\boldsymbol{n}|B)dB}. \tag{44}$$

It is worth to note that if $\sigma_{\nu,sys}$ is a constant, say $\sigma_{\nu,sys} = \sigma_{\hat{B},sys}$, where $\sigma_{\hat{B},sys}$ is the systematic error of $\hat{B}$, the maximum likelihood combined estimate of the branching ratio $B$, the Gaussian distribution $G(\nu,\sigma_{\nu,sys}^2)$ is truncated both at $\nu=0$ and at $\nu=1$. However, if one chooses $\sigma_{\nu,sys} = \tilde{\sigma}_{\nu,sys} \cdot \nu$, where $\tilde{\sigma}_{\nu,sys}$ is a constant representing for the relative error of the branching ratio $B = \nu$, we have $G(\nu,\sigma_{\nu,sys}^2) \to 0$ when $\nu \to 0$, and the truncation of $G(\nu,\sigma_{\nu,sys}^2)$ at $\nu=0$ does not appear. It is noted that the truncation at $\nu=1$ can be omitted when $\hat{B} \ll 1$.

## 5. Test with Toy Monte Carlo data

The various prescriptions described in Sections 2, 3 and 4 for the combined estimation of a branching ratio for multi-measurements are tested using toy MC data. For the two tests listed in Table 1 and Table 2, we establish two individual experiments, each of which assumes a signal pdf and a background pdf. Configurations of the tests in the tables are detailed with the notations: $L$, the joint likelihood function used in the combined estimation; $N_R$, the number of resonance decays; $\varepsilon$, the signal detection efficiency; $N_s$, the simulated number of signal events in the signal region; $f_s$, pdf for signal; $N_b$, the simulated number of background events in the signal region; $f_b$, pdf for background. $\tilde{\sigma}_{N_R}, \tilde{\sigma}_\varepsilon, \tilde{\sigma}_{N_b}$ are the relative systematic error for $N_R, \varepsilon, N_b$, respectively. $\tilde{\sigma}_{B,sys}^2 = \tilde{\sigma}_{N_R}^2 + \tilde{\sigma}_\varepsilon^2 + \tilde{\sigma}_{N_b}^2$. The tests results are given in Table 1 and Table 2 for the observed data represented by (1) same kinematic variable $m$ and same signal region, and (2) different kinematic variable $m$ and different signal region, respectively. When the systematic error of the branching ratio is taken into account in the combined estimation, a Gaussian $\pi(\nu)$ is used with standard deviation $\sigma_\nu = \sigma_{\hat{B},sys} = \hat{B} \cdot \tilde{\sigma}_{\hat{B},sys}$. Here $\tilde{\sigma}_{\hat{B},sys}$ is the relative systematic error at $\hat{B}$. In the MC simulation, the branching ratio of a mother particle decayed to a signal event is $B_{prod} = 1 \times 10^{-6}$.

Table 1 Estimate of a branching ratio for two individual experiments and the combined estimate using toy MC data. The observed spectra in the two experiments are for the same kinematic variable $m$, and have the same signal region. In the table, the binned data are formed from unbinned data using certain binning tactics. Therefore, they are actually the same except for the binning.

| Exp. | $\dfrac{N_R}{\tilde{\sigma}_{N_R}}$ | $\dfrac{\varepsilon}{\tilde{\sigma}_\varepsilon}$ | $\dfrac{N_b}{\tilde{\sigma}_{N_b}}$ | $N_s$ | $f_s$ | $f_b$ |
|---|---|---|---|---|---|---|
| 1 | $\dfrac{0.8 \times 10^8}{0.09}$ | $\dfrac{0.3}{0.15}$ | $\dfrac{2500}{0.11}$ | 24 | G(5, 0.5²) | 1st order poly. |
| 2 | $\dfrac{2.5 \times 10^8}{0.11}$ | $\dfrac{0.4}{0.12}$ | $\dfrac{2500}{0.10}$ | 100 | G(5, 1²) | 2nd order poly. |

| Spectrum format | Exp. | Joint L | pdf | without $\tilde{\sigma}_{B,sys}$ | | with $\tilde{\sigma}_{B,sys}$ | | |
|---|---|---|---|---|---|---|---|---|
| | | | | $\hat{B}$ ($10^{-6}$) | $\sigma_{\hat{B}}/\hat{B}$ | $\tilde{\sigma}_{B,sys}$ | $\hat{B}$ ($10^{-6}$) | $\sigma_{\hat{B}}/\hat{B}$ |
| Binned | 1 | | | < 2.40 | | 0.21 | < 2.44 | |
| | 2 | | | $0.97^{+0.52}_{-0.51}$ | 0.54 / 0.53 | 0.19 | $0.97^{+0.54}_{-0.53}$ | 0.56 / 0.55 |
| | 1⊕2 | Eq.(5) | Eq.(10) | $1.00^{+0.53}_{-0.52}$ | 0.53 / 0.52 | 0.14 | $1.00^{+0.55}_{-0.54}$ | 0.55 / 0.54 |
| | | Eq.(5) | Eq.(10a) | $1.01^{+0.53}_{-0.53}$ | 0.53 | 0.14 | $1.01^{+0.54}_{-0.54}$ | 0.54 |
| | | Eqs.(22,17) | | $0.96^{+0.47}_{-0.46}$ | 0.49 / 0.48 | 0.14 | $0.96^{+0.49}_{-0.48}$ | 0.52 / 0.50 |
| Unbinned | 1 | | | < 2.40 | | 0.21 | < 2.44 | |
| | 2 | | | $0.99^{+0.52}_{-0.52}$ | 0.53 | 0.19 | $0.99^{+0.54}_{-0.54}$ | 0.55 |
| | 1⊕2 | Eq.(14) | | $1.02^{+0.53}_{-0.53}$ | 0.52 | 0.14 | $1.02^{+0.54}_{-0.54}$ | 0.53 |
| | | Eq.(14a) | | $1.02^{+0.53}_{-0.53}$ | 0.52 | 0.14 | $1.02^{+0.54}_{-0.54}$ | 0.53 |
| | | Eq.(25) | | $0.97^{+0.47}_{-0.46}$ | 0.49 / 0.48 | 0.14 | $0.97^{+0.49}_{-0.48}$ | 0.51 / 0.50 |

Table 2 Estimate of a branching ratio for two individual experiments and the combined estimate using toy MC data. The observed spectra in two experiments are for different kinematic variables $m_i$, and have different signal regions. In the table, the binned data are formed from unbinned data using certain binning tactics. Therefore, they are actually the same except for the binning.

| Exp. | $\dfrac{N_R}{\tilde{\sigma}_{N_R}}$ | $\dfrac{\varepsilon}{\tilde{\sigma}_{\varepsilon}}$ | $\dfrac{N_b}{\tilde{\sigma}_{N_b}}$ | $N_s$ | $f_s$ | $f_b$ |
|---|---|---|---|---|---|---|
| 1 | $\dfrac{0.5 \times 10^8}{0.06}$ | $\dfrac{0.3}{0.14}$ | $\dfrac{2500}{0.09}$ | 24 | BW (4.6, 0.1) | 1st order poly. |
| 2 | $\dfrac{2.5 \times 10^8}{0.11}$ | $\dfrac{0.4}{0.12}$ | $\dfrac{2500}{0.10}$ | 100 | G(5, 1$^2$) | 2nd order poly. |

| Spectrum format | Exp. | Joint L | pdf | without $\tilde{\sigma}_{B,sys}$ | | with $\tilde{\sigma}_{B,sys}$ | | |
|---|---|---|---|---|---|---|---|---|
| | | | | $\hat{B}$ (10$^{-6}$) | $\sigma_{\hat{B}}/\hat{B}$ | $\tilde{\sigma}_{B,sys}$ | $\hat{B}$ (10$^{-6}$) | $\sigma_{\hat{B}}/\hat{B}$ |
| Binned | 1 | | | < 2.92 | | 0.18 | < 2.95 | |
| | 2 | | | $0.97^{+0.52}_{-0.51}$ | 0.54 / 0.53 | 0.19 | $0.97^{+0.54}_{-0.53}$ | 0.56 / 0.55 |
| | 1⊕2 | Eqs.(22,17) | Eq.(21) | $1.04^{+0.48}_{-0.47}$ | 0.47 / 0.46 | 0.13 | $1.04^{+0.50}_{-0.49}$ | 0.49 / 0.48 |
| Unbinned | 1 | | | < 2.95 | | 0.18 | < 2.99 | |
| | 2 | | | $0.99^{+0.52}_{-0.52}$ | 0.53 | 0.19 | $0.99^{+0.54}_{-0.54}$ | 0.55 |
| | 1⊕2 | Eq.(25) | | $1.06^{+0.48}_{-0.47}$ | 0.46 / 0.45 | 0.13 | $1.06^{+0.50}_{-0.49}$ | 0.48 / 0.47 |

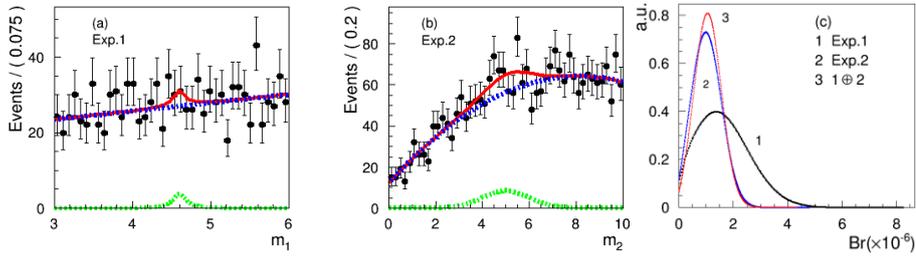

Fig. 2(color online): (a) and (b) The observed data of experiment 1 and 2 and their fitted spectra after the combined fitting is carried out for the unbinned data in Table 2. In the plots, the green, blue and red line represents the signal shape, background shape and fitted spectrum, respectively. (c) The posterior densities $h(B|\mathbf{n})$ for Exp. 1, Exp. 2 and their combination with inclusion of systematic error.

The results in Table 1 and Table 2 indicate the following features: (a) Different prescriptions of combined estimation for same multi-measurement data set give statistically coincident branching ratio values, no matter whether each individual result is presented as a central value plus error or an upper limit. (b) The accuracy of the combined branching ratio is better than each individual measurement, as expected. (c) The interval of the combined branching ratio with inclusion of systematic error is wider than that without inclusion of systematic error at the same credible level, as expected. (d) The result obtained from the unbinned likelihood function is more reliable than that from the binned likelihood, as the latter loses some measurement information.

Figures 2(a) and 2(b) show the observed data of experiment 1 and 2 and their fitted spectra after the combined fitting is carried out for the unbinned data in Table 2, while the three curves in Fig.2(c) are the posterior densities $h(B|\boldsymbol{n})$ for Exp. 1, Exp. 2 and their combination with inclusion of systematic error, respectively.

## 6. Summary

In particle physics, a decay branching ratio is often measured by different experiments, and the result of each individual measurement could be a CL=68.3% interval or a CL=90% upper limit. The combined estimate of multi-measurements will surely improve the precision of the branching ratio, however, the combined estimate with inclusion of upper limit(s) remains a difficult problem. We use the maximum likelihood method to deal with the combined estimation of multi-measurements of a branching ratio, where in each individual measurement the result can be presented as an upper limit. The joint likelihood function is constructed using the observed spectra of all experiments and the combined estimate of the branching ratio is obtained by maximizing the joint likelihood function. The Bayesian credible interval and upper limit of the combined branching ratio are given in cases both with and without inclusion of systematic error. The various prescriptions for the combined estimation of a branching ratio for multi-measurements are tested using toy MC data, which shows that different prescriptions of combined estimation for the same multi-measurements data set give statistically consistent branching ratio values, no matter whether each individual result is presented as a central value plus error or an upper limit, and the accuracy of the combined branching ratio is better than each individual measurement, as expected.